\documentstyle[aps,amssymb]{revtex}  

\def\sech\mbox{{sech\,}}    
\def\beq#1{\begin{eqnarray}\label{#1}}    
\def\eeq{\end{eqnarray}}    

\begin{document}    

\begin{flushright}
CU-TP-985
\end{flushright}   
\vskip 0.0cm

\begin{center}    
\null    

\bigskip    
   
{\LARGE \bf A Model of the Electroweak Interactions   with  
\\    
\vskip .3cm   
Invisible Higgs Particle}    
    
\bigskip    
  I.~Vitev$^{a,b}$ and  V.~Rizov$^b$   
\bigskip    
 
{\sl  $^a$ Department of Physics, Columbia University, 538 W 120-th   
Street, \\ New York, NY 10027, USA}

{\sl $^b$ Department of Theoretical Physics,  
Sofia University St.~Kliment Ohridski, \\       
boul. J. Bourchier 5, 1164 Sofia, Bulgaria}\\    
    
\medskip    
    
\end{center}

\begin{abstract}    
    
   
We propose a minimal    
unified model of the electroweak interactions without a    
Higgs particle in the final physical spectrum. This is achieved through    
adding a nonlinear constraint for    
the Higgs field in the Lagrangian in which the field's    
content is the same as in the     
Weinberg-Salam (WS) model. In the unitary gauge the generation    
of masses of the $W^{\pm}$    
and $Z$ bosons, as well as for the leptons and quarks, reproduces    
the known pattern  in the WS model. The path integral    
quantization shows that with the exception of the scalar particles'   
all other vertices known from the WS model in the unitary gauge,   
remain. A Ward identity relative to the electromagnetic    
gauge group is also derived.    
    
\end{abstract}    
    
\section{Introduction}    
    
The Weinberg-Salam model~\cite{We,Sa} uses the    
Higgs-Kibble mechanism~\cite{Hi,Ab} for the generation of    
masses for the $W^{\pm}$,$Z$    
and the spinor fields. After breaking the     
original gauge symmetry down to the electromagnetic gauge group    
one has one neutral    
Higgs boson in the physical spectrum,    
which as yet has not been observed.\footnote{After this work had been    
completed we became aware of the report at CERN about a possible   
evidence of the Higgs boson at LEP, the data, however, being   
suggestive, not conclusive. In any case we hope that the present paper   
will be of    
use for the theorists in the treatment of the Higgs and   
other elusive particles.}   
In his attempt to avoid introducing a Higgs    
field LaChapelle~\cite{La} uses the    
gauge group $\mbox{U}_c(3)\times \mbox{U}(1)\times \mbox{C}(3,1)$,    
where $U_c(3)$ is the   
color group and $\mbox{C}(3,1)$ is the conformal group acting on    
Minkowski space.    
In this model part of the gauge potentials    
(except for the photon) acquire masses but     
there remain problems with the interpretation of the other gauge fields.    
In the framework of a conformally invariant model Pawlowski and    
Raczka~\cite{Pa} also  suggest the elimination of the Higgs    
field, restricting its scalar length by a suitable    
conformal transformation.    
In the present paper we propose to eliminate the Higgs boson   
by adding a suitably chosen constraint on the scalar    
field which reduces the number of particles in the final spectrum.    
In Section II we study the classical aspects of a model of    
the electroweak interactions    
on the basis of a previously introduced gauge    
group $\mbox{MU}(2)$~\cite{R}, using a    
quadratic constraint on the Higgs field.   
As mentioned in~\cite{R}, the representations of   
$\mbox{MU}(2)$ allow one to describe fields with charges proportional    
to $1/3$ (in units of the elementary charge). This group is different from    
that proposed    
as a gauge group for the {\em standard model} in a recent    
publication by Roepstorff and Vehns ~\cite{Ro}. The latter is a subgroup $G$  
of $\mbox{SU}(5)$ and like $\mbox{MU}(2)$, $G/\mbox{SU}(3)$ also appears   
as a covering of $\mbox{U}(2)$. In Section III we    
consider the model in the unitary gauge and     
show that it reproduces all the features of the physical    
particles in the WS model     
except for the Higgs field, which is absent in the particle spectrum.    
The quantization of the model is carried out using Hamiltonian    
formulation. In Section IV we study  the proposed model as a    
system with first and second class constraints and derive    
the Hamiltonian. In Section V a Ward identity    
is derived relative to the   residual electromagnetic    
gauge group. In this paper we are not considering the     
renormalizability of the the model.    
\section{Lagrangian Formulation of the Model}    
    
Our convention for the metric in Minkowski space is     
$g_{\mu \nu} = diag (1, -1, -1, -1)$,    
the gamma matrices satisfy $\gamma_0\gamma_\mu^* \gamma_0    
= g_{\mu \mu} \gamma_\mu $     
under hermitian conjugation and the matrix     
$\gamma_5 = i\gamma_0 \gamma _1 \gamma_2 \gamma_3$ is hermitian.    
The projectors acting  in the spinor space $S$ and separating    
the right- and left-handed part of a spinor, are    
\begin{equation} \Pi_L = {1 - \gamma_5 \over 2}\;, \qquad    
\Pi_R = {1 + \gamma_5 \over 2}\;.    
\end{equation}    
The space $S$ is a direct sum     
\begin{equation}     
S = S_L \oplus S_R\;,    
\qquad S_L = \Pi_L S\;, \qquad S_R = \Pi_R S\;.    
\end{equation}    
We start with a Lagrangian that contains singlet and doublet    
states with respect to the group    
$\mbox{MU}(2) = (\mbox{R}\times \mbox{SU}(2))/3\mbox{Z}$ as suggested    
in~\cite{R},      
\begin{eqnarray}    
&{\rm left\ leptons}\, ,  \qquad&    
L^A \in C^2 \otimes S_L \subset C^2 \otimes S\;, \label{eq:1116}\\    
&{\rm left\ quarks}\, ,  \qquad&    
Q^A \in C^2 \otimes S_L \subset C^2 \otimes S\;,  \\    
&{\rm right\ leptons}\, ,  \qquad&    
R^{A}_{e} \in \Lambda^2C^2 \otimes S_R \subset \Lambda^2C^2 \otimes S\;, \\    
&{\rm right\ quarks\ of\ type\ ``p"}\, ,  \qquad&    
R^{A}_{p} \in \Lambda^2C^2 \otimes S_R \subset \Lambda^2C^2 \otimes S\;, \\    
&{\rm right\ quarks\ of\ type\ ``n"}\, ,  \qquad&    
R^{A}_{n} \in \Lambda^2C^2 \otimes S_R \subset \Lambda^2C^2 \otimes S\;, \\    
&{\rm scalar\ Higgs\ field}\, ,  \qquad&    
\phi \in C^2  \label{eq:1117}   \;,   
\end{eqnarray}    
where the index $A$ = 1,2,3 denotes the three generations of    
spinor fields. The fields (\ref{eq:1116})-(\ref{eq:1117}) transform    
under suitably chosen representations of $\mbox{MU(2)}$~\cite{R} as follows   
\begin{eqnarray}    
L^A &\ \rightarrow \ & {L'}^A = T\, {[u,A]} L^A\;, \label{eq:1119} \\    
Q^A &\ \rightarrow \ & {Q'}^A = T^{-2}\, {[u,A]} Q^A\;, \\    
R^{A}_{e} &\ \rightarrow \ & {R'}_{e}^{A} =    
{\mbox{Det}}\, {[u,A]} R^{A}_{e}\;, \\    
R^{A}_{p} &\ \rightarrow \ & {R'}_{p}^{A} =    
{\mbox{Det}}^{-{2\over 3}}\, {[u,A]} R^{A}_{p}\;, \\    
R^{A}_{n} &\ \rightarrow \ & {R'}_{n}^{A} =    
{\mbox{Det}}^{1\over 3}\, {[u,A]} R^{A}_{n}\;, \\    
\phi &\ \rightarrow \ & \phi' = T\, {[u,A]} \phi\;. \label{eq:11110}    
\end{eqnarray}    
In the notations of~\cite{R} we write    
$[u,A] \in \mbox{MU}(2)$  for the equivalence class    
$\left\{ \, \left( u + {3k\pi },e^{-{3k\pi i }} A \right)    
\ |\ k\in \mbox{Z} \, \right\},\ u \in R, \  A\in \mbox{SU(2)}$, and   
\begin{equation}   
T{[u,A]} = e^{iu}A,\qquad T^{k}{[u,A]} = e^{iu(1+{2k\over3})}A, \qquad   
\mbox{Det}^{k\over 3}{[u,A]}=e^{2iuk\over 3}.   
\end{equation}   
Since the Lie algebras of $\mbox{MU}(2)$, $\mbox{U}(2)$    
and $\mbox{SU}(2)\times \mbox{U}(1)$ are the same,   
\begin{equation}     
Lie\, \mbox{MU}(2) = \mbox{R} \oplus Lie\, \mbox{SU}(2) \;\;,     
\end{equation}      
there are three gauge potentials $A_\mu ^a$    
(a = 1,2,3) relative to the Lie algebra of $\mbox{SU}(2)$ and one, $B_\mu$,    
for $\mbox{R}$,  the respective gauge coupling parameters    
being $g$ and $g'$.    
A set of four generators for $\mbox{MU}(2)$     
is given by     
\begin{equation}     
X^a = \left( 0,{\sigma ^a \over 2} \right)\, , \ a=1,2,3 \;,\qquad      
{\rm and}  \qquad X = \left( -{1\over 2},0 \right)\; ,     
\end{equation}     
where $\sigma^a$ are the Pauli matrices. Note that it is specific   
for this model that the actions of    
$\mbox{MU}(2)$ on the left    
leptons $L^{A}$ and on the Higgs field $\phi$ coincide,    
so that the covariant derivative of $\phi$     
reads    
\begin{equation}    
{\cal D}_{\mu} \phi = \left( {\partial}_{\mu} -    
igA^{a}_{\mu}{{\sigma}^a \over 2} +    
ig' B_{\mu}{I\over 2} \right) \phi \;.    
\end{equation}    
The remaining covariant     
derivatives are the same as in the WS model. The Yang-Mills Lagrangian    
\begin{equation}    
{\cal L}_{\mbox {\scriptsize YM}} =    
- {1\over 4}  F^{a}_{\mu \nu} F^{\mu \nu ,a}    
- {1\over 4}  B_{\mu \nu}  B^{\mu \nu}   
\end{equation}    
contains    
\begin{equation}    
F^{a}_{\mu \nu} = \partial_{\mu}A^{a}_{\nu} -    
\partial_{\nu}A^{a}_{\mu} + g \varepsilon^{abc}A^{b}_{\mu}A^{c}_{\nu}\;,    
\qquad   
B_{\mu \nu} = \partial_{\mu}B_{\nu} -    
\partial_{\nu}B_{\mu}\; .    
\end{equation}    
In order to write the scalar-spinor interaction    
term in an $\mbox{MU}(2)$ invariant form we    
note the following:   
let $e_1,e_2$ be a basis in ${\Bbb C}^2$, $\phi \in {\Bbb C}^2$, $u \in     
{\Bbb C}^2 \otimes S $, and $G$ be the standard hermitian metric    
in ${\Bbb C}^2$.    
Writing $\phi = \phi_1e_1 + \phi_2e_2$, $u = u_1e_1 + u_1e_2$    
one has     
$\phi\wedge u = (\phi_1u_2 - \phi_2u_1)e_1\wedge e_2    
\in \Lambda ^2 {\Bbb C}^2\otimes S$.    
Using $G$ and Dirac conjugation we define a bilinear form    
in $\Lambda ^2 {\Bbb C}^2\otimes S$ denoted by     
$\langle \ , \ \rangle_G$ such that    
\begin{equation}    
 \langle \psi \, e_1\wedge e_2 ,\theta \, e_1\wedge e_2\rangle_G    
= \bar{\psi}\theta \;.   
\end{equation}    
Let $\tilde{u} = i\sigma^2 u^*$ where $u^*$ is the complex     
conjugate of $u$.   
We will further use the representations of $\mbox{MU}(2)$    
on $\Lambda ^2 {\Bbb C}^2$ defined    
by $\mbox{Det}^{k\over 3}{[u,A]}=e^{2iuk\over 3}$.    
We may write now an $\mbox{MU}(2)$ invariant term involving    
the spinor fields and the  Higgs field, namely    
\begin{equation}    
{\cal L}_{\rm Yuk} = - \left[\,   
K^{e}_{AB} {\langle\phi \wedge L^A ,R^{B}_e\rangle}_G +    
K^{p}_{AB} {\langle\tilde{\phi} \wedge Q^A ,R^{B}_p\rangle}_G +    
K^{n}_{AB} {\langle\phi \wedge Q^A ,R^{B}_n\rangle}_G + h.c.  
\, \right]\;,   
\end{equation}    
summation over the repeated indices $A,B$ is assumed.   
Here $K^{e}_{AB}$ are the matrix elements of a    
$3\times 3$ real diagonal matrix, while    
$K^{p}_{AB}$ and $K^{n}_{AB}$ are the matrix elements    
of $3\times 3$ complex invertible    
matrices. Note that the transformation property of $\phi$  
yields a form of ${\cal L}_{\rm Yuk}$ which contains a minor difference   
as compared to the explicit form of the    
Yukawa term in the WS model.    
The pure Higgs field term reads    
\begin{equation}\label{eq:11101}    
{\cal L}_{\phi} = {({\cal D}_{\mu} \phi)}^* ({\cal D}^{\mu} \phi) +    
c(x)(\phi ^* \phi - a^2)\;, \qquad    
a=const.\, , \quad a>0\;.    
\end{equation}    
Instead of the standard potential    
 $V(\phi) = - \mu^2 \phi^* \phi +\lambda (\phi^* \phi)^2 $    
it contains a nonlinear constraint fixing the squared norm of    
the field $\phi$,     
$||\phi||^2 = a^2$.   
The real field $c(x)$ is the corresponding Lagrange multiplier.    
Considering the equations of motion for $\phi(x)$ and $c(x)$,    
we find    
\begin{eqnarray}    
&&{\cal D}_{\mu} {\cal D}^{\mu} \phi -    
c(x)\phi - V = 0\; ,\label{eq:11113} \\    
&&\phi^*{\cal D}^{*}_{\mu} {\cal D}^{*\mu}  -    
c(x)\phi^* - V^* = 0\; ,\label{eq:11114} \\    
&&\phi^* \phi - a^2 = 0\; . \label{eq:11115}    
\end{eqnarray}    
${\cal D}^{*}_{\mu}$ implies    
hermitian conjugation  of the covariant derivative as well as action    
of the differential operator to the left.   
The explicit form of $V \in {\Bbb C}^2$ is    
\begin{equation}    
V={\partial {\cal L}_{\rm Yuk} \over {\partial \phi^*}} =    
\left( \begin{array}{c}    
-K^{e}_{AB}{({\bar{L}}^A)}_2 R_{e}^{B}    
-K^{*p}_{AB}{\bar{R}}^{A}_{p} {(Q^{B})}_1    
-K^{n}_{AB}{({\bar{Q}}^A)}_2 R_{n}^{B} \\    
\\    
K^{e}_{AB}{({\bar{L}}^A)}_1 R_{e}^{B}    
-K^{*p}_{AB}{\bar{R}}^{A}_{p} {(Q^{B})}_2    
+K^{n}_{AB}{({\bar{Q}}^A)}_1 R_{n}^{B}    
\end{array} \right)    
\end{equation}    
and $V^*$ is the hermitian conjugate of $V$.   
Expressing $c(x)$    
from (\ref{eq:11113})-(\ref{eq:11115})    
\begin{eqnarray}    
c(x) &=& {1\over a^2}\left[\phi^* ({\cal D}_{\mu}    
{\cal D}^{\mu}\phi ) - \phi^* V\right]\;, \label{eq:111001}\\    
c(x) &=& {1\over a^2}\left[(\phi^* {\cal D}^{*}_{\mu}    
{\cal D}^{*\mu})\phi  - V^* \phi\right] \;, \label{eq:111002}    
\end{eqnarray}    
we eliminate the Lagrange multiplier and obtain   
\begin{eqnarray}    
&&{\cal D}_{\mu} {\cal D}^{\mu} \phi -    
{1\over a^2}\left[\phi^* ({\cal D}_{\mu}    
{\cal D}^{\mu}\phi ) - \phi^* V\right]    
\phi - V = 0\;,\label{eq:11116} \\    
&&\phi^*{\cal D}^{*}_{\mu} {\cal D}^{*\mu}  -    
{1\over a^2}\phi^* \left[(\phi^* {\cal D}^{*}_{\mu}    
{\cal D}^{*\mu})\phi  - V^* \phi\right]    
- V^* = 0\;,\label{eq:11117} \\    
&&\phi^* ({\cal D}_{\mu}    
{\cal D}^{\mu}\phi ) - \phi^* V =    
(\phi^* {\cal D}^{*}_{\mu}    
{\cal D}^{*\mu})\phi  - V^* \phi\;. \label{eq:11118}    
\end{eqnarray}    
\section{The Unitary Gauge}    
    
The unitary gauge is the gauge in which the scalar Higgs field    
has one constant component. Due to the nonlinear constraint     
(\ref{eq:11115})  the  other    
component is also fixed    
\begin{equation}\label{eq:11119}    
\phi_0 (x) = \left( \begin{array}{c} a \\ 0 \end{array}    
\right)\;.    
\end{equation}    

The electromagnetic subgroup $\mbox{MU}_{\mbox {em}}(1)$ is the little group for    
$\phi_0$, i.e.     
\begin{equation}    
T\ { \left[-{\alpha \over 2}, \left( \begin{array}{cc}    
\scriptsize    
e^{i\alpha \over 2} &\scriptsize 0  \\    
\scriptsize 0 & \scriptsize e^{i\alpha \over 2}, \end{array}    
\right) \right] } \phi_0 = \phi_0,  \qquad \alpha\in\mbox{R}\;.   
\end{equation}    
For each point $x$ we may choose an element $[u,A] \in \mbox{MU}(2)$,    
so that $T\, {[u,A]}$ transforms $\phi$ to the form (\ref{eq:11119}),    
namely    
\begin{eqnarray}    
[u,A] = \left[ 0 , {1\over a}\left( \begin{array}{cc}    
\varphi^{*}_{1} & \varphi^{*}_{2}   \\    
- \varphi_{2} & \varphi_{1}  \end{array}    
\right) \right] \; , \qquad    
T\ {[u,A]} =  {1\over a}\left( \begin{array}{cc}    
\varphi^{*}_{1} & \varphi^{*}_{2}   \nonumber \\    
- \varphi_{2} & \varphi_{1}  \end{array}    
\right)\; , \\    
\phi = \left( \begin{array}{c} \phi_{1} \\ \phi_{2}    
\end{array} \right)\;,  \qquad \phi^* \phi = {||\phi||}^2 =    
{|\phi_{1}|}^2 + {|\phi_{2}|}^2 = a^2\;.   
\end{eqnarray}    
Following the standard pattern we define the physical fields $W^{\pm}_{\mu},\ Z_{\mu}$    
and $A_{\mu}$ as unitary linear combinations of  the original gauge    
fields. The four fields with definite electric charge are   
\begin{equation}\label{eq:11120}    
\left( \begin{array}{c} W^{+}_{\mu} \\ W^{-}_{\mu}    
\end{array} \right) = \left( \begin{array}{cc}    
{1\over \sqrt{2}} & -{i\over \sqrt{2}}     \\    
{1\over \sqrt{2}} & {i\over \sqrt{2}}  \end{array}    
\right) \left( \begin{array}{c} A^{1}_{\mu} \\ A^{2}_{\mu}    
\end{array} \right)\;, \qquad    
\left( \begin{array}{c} Z_{\mu} \\ A_{\mu}    
\end{array} \right) = \left( \begin{array}{cc}    
{g\over \sqrt{g^2 + {g'}^2}} & -{g'\over \sqrt{g^2 + {g'}^2}} \\    
{g'\over \sqrt{g^2 + {g'}^2}} & {g\over \sqrt{g^2 + {g'}^2}}    
\end{array}  \right) \left( \begin{array}{c} A^{3}_{\mu} \\    
B_{\mu} \end{array} \right)\;.    
\end{equation}    
Indeed, the general gauge transformation     
\begin{equation}\label{eq:11100}    
{A'}^{a}_{\mu} T^a = T(g) (A^{a}_{\mu} T^a){T(g)}^{-1}    
- {i\over g}({\partial}_{\mu}T(g)) {T(g)}^{-1},    
\end{equation}    
applied for an element    
from the electromagnetic subgroup $\mbox{MU}_{\mbox {em}}(1)$, gives    
for the fields (\ref{eq:11120})    
\begin{eqnarray}    
&&{W'}^{+}_{\mu} = e^{i\alpha}  W^{+}_{\mu}\; ,   \qquad    
{W'}^{-}_{\mu} = e^{-i\alpha} W^{-}_{\mu}\;, \\    
&&{Z'}_{\mu} = Z_{\mu}\;,   \qquad \quad {A'}_{\mu} =    
A_{\mu} +{1\over e} {\partial}_{\mu} \alpha\;.    
\end{eqnarray}    
The elementary electric charge is identified as    
$e={gg'\over {\sqrt {g^2 +{g'}^2}}}$.    
Therefore, in the unitary gauge after    
breaking the symmetry down to the $\mbox{MU}_{\mbox {em}}(1)$ subgroup,    
we recognize with no surprise that   
$W^{\pm}_{\mu}$ are charged vector fields, $Z_{\mu}$ is a    
neutral vector field and $A_{\mu}$ is the remaining gauge    
potential (for the residual invariance group), identified    
with the electromagnetic field.    
    
If we go back to the Higgs field term (\ref{eq:11101}), we see    
that in the unitary gauge it only contributes to the masses    
of the vector fields    
\begin{eqnarray}    
&&{\cal L}_{\phi} = {g^2 a^2 \over 2} W^{+}_{\mu} W^{-\mu} +    
{(g^2 + {g'}^2) a^2 \over 4} Z_{\mu} Z^{\mu}\;, \\    
&&M^{2}_{W} = {g^2 a^2 \over 2} \;, \qquad    
M^{2}_{Z} = {(g^2 + {g'}^2) a^2 \over 2} \;. \label{eq:11123}    
\end{eqnarray}    
The field $c(x)$ has been excluded and does not appear in the unitary gauge.    
As in the standard scheme, in the unitary gauge the lepton fields    
are identified with the physical leptons, whereas the quark    
fields appear as linear combinations of quarks with definite    
current masses    
\begin{equation}    
L^A = \left( \begin{array}{c} \nu^{A}_{L} \\[.5ex] e^{A}_{L} \end{array}    
\right)\;, \quad R^{A}_{e} = e^{A}_{R}\;, \qquad    
\begin{array}{l} \nu^A = \nu_e,\nu_{\mu},\nu_{\tau} \\[1ex]    
e^A = e,\mu,\tau \end{array} \;,    
\end{equation}    
\vskip -10pt    
\begin{equation}    
Q^A = \left( \begin{array}{c} {p'}^{A}_{L} \\[.5ex] {n'}^{A}_{L}    
\end{array} \right)\;, \quad \begin{array}{l}    
R^{A}_{p} = {p'}^{A}_{R} \\[1ex] R^{A}_{n} = {n'}^{A}_{R}    
\end{array} \;,\qquad    
\begin{array}{l} {p'}^A = u',c',t' \\[1ex]    
{n'}^A = d',s',b' \end{array} \;.   
\end{equation}    

The scalar-spinor term in the Lagrangian after breaking the symmetry    
acquires the form    
\begin{equation}\label{eq:11128}    
{\cal L}_{\rm Yuk} = -\left[ \,    
{\tilde {M}}^{e}_{AB}{\bar{e}}^{A}_{R} e^{B}_{L} +    
{\tilde {M}}^{p}_{AB}{\bar{p'}}^{A}_{R} {p'}^{B}_{L} +    
{\tilde {M}}^{n}_{AB}{\bar{n'}}^{A}_{R} {n'}^{B}_{L} +    
\mbox {h.c.} \, \right]\;.    
\end{equation}    
The mass matrices for leptons and quarks have matrix elements    
${\tilde{M}}^{e}_{AB}=aK^{e}_{AB},\ {\tilde{M}}^{p}_{AB}=aK^{p}_{AB},\    
{\tilde{M}}^{n}_{AB}=aK^{n}_{AB}$. The neutrinos are strictly massless    
and the quark mass matrices are in general non-diagonal.    
The procedure of their  diagonalization goes in the same way as    
in the WS model yielding  the quark mass eigenstates    
$p^A$ and $n^A$~\cite{Lee}.    

The Lagrangian of the proposed model in the unitary gauge can    
be written in a form, suitable for determining the canonical    
momenta of the fields and the primary constraints for a further    
 transition to    
Hamiltonian formalism and quantization by path integrals. The term    
in the Lagrangian, associated with the massive vector fields and    
the electromagnetic gauge potential, is    
\begin{eqnarray}\label{eq:10000}    
{\cal L}_{\mbox {vec.}} =    
&-&{1\over 4}(\partial_{\mu}A_{\nu} - \partial_{\nu}A_{\mu})    
(\partial^{\mu}A^{\nu} - \partial^{\nu}A^{\mu})     
-{1\over 4}(\partial_{\mu}Z_{\nu} - \partial_{\nu}Z_{\mu})    
(\partial^{\mu}Z^{\nu} - \partial^{\nu}Z^{\mu}) \nonumber \\    
 &+& {1\over 2} M^{2}_{Z}Z_{\mu} Z^{\mu}    
-{1\over 2}(\partial_{\mu}W^{+}_{\nu} - \partial_{\nu}W^{+}_{\mu})    
(\partial^{\mu}W^{-\nu} - \partial^{\nu}W^{-\mu})    
  +  M^{2}_{W} W^{+}_{\mu} W^{-\mu} \nonumber \\    
 &+& {igg'\over {\sqrt{ g^2 +{g'}^2}}}\left[ (\partial_{\mu}A_{\nu} -    
\partial_{\nu}A_{\mu})W^{+\mu} W^{-\nu} +    
(\partial_{\mu}W^{+}_{\nu} - \partial_{\nu}W^{+}_{\mu})    
W^{-\mu} A^{\nu} - (\partial_{\mu}W^{-}_{\nu} -    
\partial_{\nu}W^{-}_{\mu})A^{\nu}W^{+\mu} \right]  \nonumber \\    
 &+& {ig^2\over {\sqrt {g^2 +{g'}^2}}}\left[ (\partial_{\mu}Z_{\nu} -    
\partial_{\nu}Z_{\mu})W^{+\mu} W^{-\nu} +    
(\partial_{\mu}W^{+}_{\nu} - \partial_{\nu}W^{+}_{\mu})    
W^{-\mu} Z^{\nu} - (\partial_{\mu}W^{-}_{\nu} -    
\partial_{\nu}W^{-}_{\mu})Z^{\nu}W^{+\mu} \right]   \nonumber \\    
 &-& {g^2\over 2}(W^{+}_{\mu} W^{-\mu}W^{+}_{\nu} W^{-\nu} -    
W^{+}_{\mu} W^{+\mu}W^{-}_{\nu} W^{-\nu}) -    
{g^2{g'}^2\over  g^2 +{g'}^2} (W^{+}_{\mu} W^{-\mu}A_{\nu}A^{\nu} -    
W^{+}_{\mu} A^{\mu}W^{-}_{\nu}A^{\nu}) \nonumber \\    
 &-& {g^4\over  g^2 +{g'}^2} (W^{+}_{\mu} W^{-\mu}Z_{\nu}Z^{\nu} -    
W^{+}_{\mu} Z^{\mu}W^{-}_{\nu}Z^{\nu}) -    
{g^3g'\over  g^2 +{g'}^2} (2W^{+}_{\mu} W^{-\mu}A_{\nu}Z^{\nu} -    
W^{+}_{\mu} A^{\mu}W^{-}_{\nu}Z^{\nu} -    
W^{+}_{\mu} Z^{\mu}W^{-}_{\nu}A^{\nu})\;.  \nonumber \\   
\end{eqnarray}    
The Lagrangian, describing a free spinor theory with    
physical fermions, is    
\begin{equation}    
{\cal L}^{(0)}_{\mbox {f}} = {\bar{e}}^{A}    
(i\gamma^{\mu} \partial_{\mu} - m^{e}_{A})e^A +    
{\bar{p}}^{A}(i\gamma^{\mu} \partial_{\mu} - m^{p}_{A})p^A +    
{\bar{n}}^{A}(i\gamma^{\mu} \partial_{\mu} - m^{n}_{A})n^A +    
{\bar{\nu}}^{A}i\gamma^{\mu} \partial_{\mu}\nu^A\;.    
\end{equation}    
Here $m^{e}_{A}$, $m^{p}_{A}$ and $m^{n}_{A}$ denote    
the masses of the three   charged leptons    
and quarks of type ``$p$" and ``$n$" respectively.    
It remains to list the electromagnetic current    
part, the weak neutral current part and the charged    
current term of the Lagrangian    
\begin{equation}    
{\cal L}_{\mbox {em.c.}} =  {gg'\over {\sqrt {g^2 +{g'}^2}}} \left[    
-{\bar{e}}^{A}\gamma^{\mu} e^A + {2\over 3}    
{\bar{p}}^{A}\gamma^{\mu} p^A - {1\over 3}    
{\bar{n}}^{A}\gamma^{\mu} n^A \right]A_{\mu} \;.    
\end{equation}    
\begin{eqnarray}    
{\cal L}_{\mbox {n.c.}} = && {1\over {2\sqrt {g^2 +{g'}^2}}}\left[    
-(g^2 - {g'}^2 ){\bar{e}}^{A}_{L}\gamma^{\mu} e^{A}_{L}    
+2{g'}^2{\bar{e}}^{A}_{R}\gamma^{\mu} e^{A}_{R}    
+(g^2 + {g'}^2 ){\bar{\nu}}^{A}\gamma^{\mu}    
\nu^{A}\right]Z_{\mu}  \nonumber \\    
&+&{1\over {2\sqrt{ g^2 +{g'}^2}}}\left[   
(g^2 - {{g'}^2\over 3} ){\bar{p}}^{A}_{L}\gamma^{\mu} p^{A}_{L}    
-{4{g'}^2\over 3}{\bar{p}}^{A}_{R}\gamma^{\mu} p^{A}_{R}    
-(g^2 + {{g'}^2\over 3} ){\bar{n}}^{A}_{L}\gamma^{\mu} n^{A}_{L}    
+{2{g'}^2\over 3}{\bar{n}}^{A}_{R}\gamma^{\mu} n^{A}_{R}    
\right]   Z_{\mu}\;.    
\end{eqnarray}    
\begin{equation}\label{eq:10001}    
{\cal L}_{\mbox {c.c.}} =  {g\over {\sqrt 2}}\left[    
 ({\bar{\nu}}^{A}_{L}\gamma^{\mu} e^{A}_{L} +    
{\bar{p}}^{A}_{L}\gamma^{\mu} U_{AB}n^{B}_{L})W^{+}_{\mu}    
+ \mbox{h.c.} \right]\;,    
\end{equation}    
where $U$ is the Kobayashi-Maskawa matrix.   
%
Finally we go back to the equations (\ref{eq:11116})-(\ref{eq:11118}).    
If we take into account the explicit form of the vectors $V$ and    
$V^*$ in the unitary gauge, we find    
\begin{eqnarray}    
&&\partial_{\mu}W^{+\mu} +  {i{g'}^2\over {\sqrt {g^2 +{g'}^2}}}    
Z_{\mu}W^{+\mu}- {igg'\over {\sqrt {g^2 +{g'}^2}}}A_{\mu}W^{+\mu}   
\nonumber \\    
&&\qquad\qquad \qquad   
+{i\sqrt{2}\over ga^2} (m^{e}_{A}{\bar{e}}^{A}_{R}\nu^{A} +    
m^{n}_{A}{\bar{n}}^{A}_{R}U^{*}_{AB}p^{B}_{L} -    
m^{p}_{B}{\bar{n}}^{A}_{L}U^{*}_{AB}p^{B}_{R}) = 0\;,  \label{qq} \\    
&&\partial_{\mu}W^{-\mu} -  {i{g'}^2\over {\sqrt {g^2 +{g'}^2}}}    
Z_{\mu}W^{-\mu}+ {igg'\over {\sqrt {g^2 +{g'}^2}}}A_{\mu}W^{-\mu}   
\nonumber \\     
&&\qquad \qquad \qquad   
-{i\sqrt{2}\over ga^2} (m^{e}_{A}{\bar{\nu}}^{A}e^{A}_{R} +    
m^{n}_{B}{\bar{p}}^{A}_{L}U_{AB}n^{B}_{R} -    
m^{p}_{A}{\bar{p}}^{A}_{R}U_{AB}n^{B}_{L}) = 0\;, \\    
&&\partial_{\mu}Z^{\mu} -  {i\over {a^2\sqrt {g^2 +{g'}^2}}}    
\left[m^{e}_{A}({\bar{e}}^{A}_{R}e^{A}_{L} -{\bar{e}}^{A}_{L}e^{A}_{R})    
\right. \nonumber \\    
&&\left. \qquad \qquad\qquad   
-m^{p}_{A}({\bar{p}}^{A}_{R}p^{A}_{L}-    
{\bar{p}}^{A}_{L}p^{A}_{R})    
+m^{n}_{A}({\bar{n}}^{A}_{R}n^{A}_{L} -{\bar{n}}^{A}_{L}n^{A}_{R})\right]    
=0\;. \label{qqq}    
\end{eqnarray}    
Equations~(\ref{qq})-(\ref{qqq}) appear in the unitary gauge    
as part of the equations of motion    
for the $W^{\pm}$ and $Z$ bosons. They signify     
the fact that a massive vector field has three    
physical components and are analogous to $\partial_{\mu}U^{\mu}=0$    
in Proka's theory of a free vector field $U^{\mu}$.    
\section{Hamiltonian Formalism}    
    
For the transition to a Hamiltonian form and    
a subsequent quantization of the model    
we will use the approach and terminology of \cite{Sen,Git,Sl}.    
Following \cite{Git},  we will    
treat the fields in the model as elements of    
a Berezin algebra.    
The time component of the electromagnetic gauge potential    
$A_0$ will be considered as a Lagrange multiplier.    
    
Given the Lagrangian (\ref{eq:10000})-(\ref{eq:10001})    
we find the conjugate momenta for the spinor fields   
\begin{eqnarray}    
&&\Pi_{e^A} = {\partial _R {\cal{L}}\over    
\partial \mathaccent 95{e}^A}=i\bar{e}^A\gamma^0\;, \qquad    
\Pi_{\bar{e}^A} = {\partial_R {\cal{L}}\over    
\partial {\mathaccent 95{\bar{e}}}^A}=0\;, \label{eq:2225}\\    
&&\Pi_{\nu^A} = {\partial_R {\cal{L}}\over    
\partial {\mathaccent 95{\nu}}^A}=i\bar{\nu}^A\gamma^0\;, \qquad    
\Pi_{\bar{\nu}^A} = {\partial_R {\cal{L}}\over    
\partial {\mathaccent 95{\bar{\nu}}}^A}=0\;, \\    
&&\Pi_{p^A} = {\partial_R {\cal{L}}\over    
\partial {\mathaccent 95{p}}^A}=i\bar{p}^A\gamma^0\;, \qquad    
\Pi_{\bar{p}^A} = {\partial_R {\cal{L}}\over    
\partial {\mathaccent 95{\bar{p}}}^A}=0\;, \\    
&&\Pi_{n^A} = {\partial_R {\cal{L}}\over    
\partial {\mathaccent 95n}^A}=i\bar{n}^A\gamma^0\;, \qquad    
\Pi_{\bar{n}^A} = {\partial_R {\cal{L}}\over    
\partial {\mathaccent 95{\bar{n}}}^A}=0\;.    
\end{eqnarray}    
Here    
${\partial _R {\cal{L}}\over \partial \mathaccent 95{e}^A}$,   
etc..., stand for ``right" differentiation. The canonical    
momenta for the massive vector fields and    
the electromagnetic gauge potential are   
\begin{eqnarray}    
&&{\Pi}_{0}^{Z} = {{\partial}_R {\cal L} \over    
\partial \mathaccent 95{Z}^0}=0\;, \\    
&&{\Pi}_{i}^{Z} = {{\partial}_R {\cal L} \over    
\partial \mathaccent 95{Z}^i}=  \mathaccent 95{Z}^i+    
\partial_i Z_0 +  {ig^2\over {\sqrt{ g^2 +{g'}^2}}}    
\left( W^{+}_{0} W^{-}_{i} - W^{+}_{i} W^{-}_{0}\right)\;, \\    
&&{\Pi}_{0}^{W^+} = {{\partial}_R {\cal L} \over    
\partial \mathaccent 95{W}^{+0}}=0\;, \\    
&&{\Pi}_{i}^{W^+} = {{\partial}_R {\cal L} \over    
\partial \mathaccent 95{W}^{+i}}=  \mathaccent 95{W}^{-i}+    
\partial_i W_{0}^{-} +  {igg'\over {\sqrt{ g^2 +{g'}^2}}}    
\left( W^{-}_{0} A_{i} - A_0 W^{-}_{i} \right) +    
  {ig^2 \over {\sqrt{ g^2 +{g'}^2}}}    
\left( W^{-}_{0} Z_{i} - Z_0 W^{-}_{i} \right)\;,\\    
&&{\Pi}_{0}^{W^-} = {{\partial}_R {\cal L} \over    
\partial \mathaccent 95{W}^{-0}}=0\;, \\    
&&{\Pi}_{i}^{W^-} = {{\partial}_R {\cal L} \over    
\partial \mathaccent 95{W}^{-i}}=  \mathaccent 95{W}^{+i}+    
\partial_i W_{0}^{+} +  {igg'\over {\sqrt{ g^2 +{g'}^2}}}    
\left( A_0 W^{+}_{i} -  W^{+}_{0}A_i \right) +    
  {ig^2 \over {\sqrt{ g^2 +{g'}^2}}}    
\left( Z_0 W^{+}_{i} -  W^{+}_{0} Z_i \right)\;,\\    
&&{\Pi}_{i}^{A} = {{\partial}_R {\cal L} \over    
\partial \mathaccent 95{A}^i}=  \mathaccent 95{A}^i+    
\partial_i A_0 +  {igg'\over {\sqrt{ g^2 +{g'}^2}}}    
\left( W^{+}_{0} W^{-}_{i} - W^{+}_{i} W^{-}_{0}\right)\;. \label{eq:2226}    
\end{eqnarray}    
The corresponding velocities that can be expressed from here are    
$\mathaccent 95{W}^{\pm i},\  \mathaccent 95{Z}^i$    
and $\mathaccent 95{A}^i$.    
The remaining equations from    
(\ref{eq:2225})-(\ref{eq:2226}) and the part of the Lagrangian    
that multiplies $A^0$ define the primary constraints in the model    
\begin{eqnarray}    
&&\phi^{(1)}_{e^A} = {\Pi}_{e^A} -    
i\bar{e}^A\gamma^0\;, \qquad \quad    
\phi^{(1)}_{\bar{e}^A}={\Pi}_{\bar{e}^A} \;, \\    
&&\phi^{(1)}_{\nu^A} = {\Pi}_{\nu^A} -    
i\bar{\nu}^A\gamma^0\;, \qquad \quad    
\phi^{(1)}_{\bar{\nu}^A}={\Pi}_{\bar{\nu}^A} \;, \\    
&&\phi^{(1)}_{p^A} = {\Pi}_{p^A} -    
i\bar{p}^A\gamma^0\;, \qquad \quad    
\phi^{(1)}_{\bar{p}^A}={\Pi}_{\bar{p}^A} \;, \\    
&&\phi^{(1)}_{n^A} = {\Pi}_{n^A} -    
i\bar{n}^A\gamma^0, \qquad \quad    
\phi^{(1)}_{\bar{n}^A}={\Pi}_{\bar{n}^A} \;.    
\end{eqnarray}    
For the massive vector mesons and the electromagnetic    
gauge potential one gets  the primary constraints  
\begin{eqnarray}    
&&\phi^{(1)}_{Z} = {\Pi}_{0}^{Z}\;, \qquad    
\phi^{(1)}_{W^+}= {\Pi}_{0}^{W^+}\;, \qquad    
\phi^{(1)}_{W^-}= {\Pi}_{0}^{W^-}\;, \nonumber \\ &&\nonumber \\    
&&\phi^{(1)}_{A}=\partial_i\Pi_{i}^{A} +    
{igg'\over {\sqrt{ g^2 +{g'}^2}}}    
\left( \Pi^{W^-}_{i} W^{-}_{i} - \Pi^{W^+}_{i}W^{+}_{i}\right) \nonumber\\    
&&\qquad\qquad \qquad  - {gg'\over {\sqrt {g^2 +{g'}^2}}} \left(    
-{\bar{e}}^{A}\gamma^{0} e^A + {2\over 3}    
{\bar{p}}^{A}\gamma^{0} p^A - {1\over 3}    
{\bar{n}}^{A}\gamma^{0} n^A \right)\;.  
\end{eqnarray}    
    
With the help of the explicitly solved velocities one finds the    
Hamiltonian of the system    
\begin{eqnarray}    
{\cal H} = &{1\over 2}&\Pi^{A}_{i}\Pi^{A}_{i} -    
{igg'\over {\sqrt {g^2 +{g'}^2}}}\Pi^{A}_{i}    
\left( W^{+}_{0} W^{-}_{i} - W^{+}_{i} W^{-}_{0}\right)+    
{1\over 2}\Pi^{Z}_{i}\Pi^{Z}_{i} -    
{ig^2\over {\sqrt {g^2 +{g'}^2}}}\Pi^{Z}_{i}    
\left( W^{+}_{0} W^{-}_{i} - W^{+}_{i} W^{-}_{0}\right) \nonumber \\[.5ex]      
&-&\Pi^{Z}_{i}\partial_{i}Z_0 +    
\Pi^{W^+}_{i}\Pi^{W-}_{i} +    
{igg'\over {\sqrt{ g^2 +{g'}^2}}}A_i    
 (\Pi^{W^+}_{i} W^{+}_{0} - W^{-}_{0}\Pi^{W^-}_{i})+    
{ig^2\over {\sqrt{ g^2 +{g'}^2}}}    
\left( W^{+}_{0} Z_i - Z_0 W^{+}_{i} \right)\Pi^{W^+}_{i} \nonumber \\[.5ex]      
&-&{ig^2\over {\sqrt{ g^2 +{g'}^2}}}    
\Pi^{W^-}_{i} \left( W^-_{0} Z_i - Z_0 W^{-}_{i} \right)-    
\Pi^{W^-}_{i}\partial_{i}W^{-}_{0} -    
\Pi^{W^+}_{i}\partial_{i}W^{+}_{0} +    
{1\over 2}M^{2}_{Z}{Z_0}{Z_0}+{1\over 2}M^{2}_{Z}{Z_i}{Z_i}    
+M^{2}_{W}W^{+}_{0}W^{-}_{0} \nonumber\\ [.5ex]     
&+&M^{2}_{W}W^{+}_{i}W^{-}_{i}+    
{1\over 4}(\partial_{i}A_{k} - \partial_{k}A_{i})^2   
+{1\over 4}(\partial_{i}Z_{k} - \partial_{k}Z_{i})^2+   
{1\over 2}(\partial_{i}W^{+}_{k} - \partial_{k}W^{+}_{i})    
(\partial_{i}W^{-}_{k} - \partial_{k}W^{-}_{i}) \nonumber \\[.5ex]      
& -& {igg'\over {\sqrt{ g^2 +{g'}^2}}}\left[ (\partial_{i}A_{k} -    
\partial_{k}A_{i})W^{+}_{i} W^{-}_{k} +    
(\partial_{i}W^{+}_{k} - \partial_{k}W^{+}_{i})    
W^{-}_{i} A_k + (\partial_{i}W^{-}_{k} -    
\partial_{k}W^{-}_{i})A_iW^{+}_{k} \right] \nonumber \\ [.5ex]     
 &-& {ig^2\over {\sqrt {g^2 +{g'}^2}}}\left[ (\partial_{i}Z_{k} -    
\partial_{k}Z_{i})W^{+}_{i} W^{-}_{k} +    
(\partial_{i}W^{+}_{k} - \partial_{k}W^{+}_{i})    
W^{-}_{i} Z_k + (\partial_{i}W^{-}_{k} -    
\partial_{k}W^{-}_{i})Z_iW^{+}_{k} \right] \nonumber \\[.5ex]      
 &+& {g^2\over 2}(W^{+}_{i} W^{-}_{i}W^{+}_{k} W^{-}_{k} -    
W^{+}_{i} W^{+}_{i}W^{-}_{k} W^{-}_{k}) +    
{g^2{g'}^2\over  g^2 +{g'}^2} (W^{+}_{i} W^{-}_{i}A_{k}A_{k} -    
W^{+}_{i} A_iW^{-}_{k}A_k) \nonumber \\[.5ex]      
 &+& {g^4\over  g^2 +{g'}^2} (W^{+}_{i} W^{-}_{i}Z_{k}Z_k -    
W^{+}_{i} Z_iW^{-}_{k}Z_k) +    
{g^3g'\over  g^2 +{g'}^2} (2W^{+}_{i} W^{-}_{i}A_{k}Z_k -    
W^{+}_{i} A_iW^{-}_{k}Z_k -    
W^{+}_{i} Z_iW^{-}_{k}A_k)  \nonumber \\[.5ex]    
&+& {\bar{e}}^{A}    
(i\gamma_k \partial_{k} - m^{e}_{A})e^A +    
{\bar{p}}^{A}(i\gamma_k \partial_{k} - m^{p}_{A})p^A +    
{\bar{n}}^{A}(i\gamma_k \partial_{k} - m^{n}_{A})n^A +    
{\bar{\nu}}^{A}i\gamma_k \partial_{k}\nu^A \nonumber \\[.5ex]     
&-& {gg'\over {\sqrt {g^2 +{g'}^2}}} \left[    
-{\bar{e}}^{A}\gamma_k e^A + {2\over 3}    
{\bar{p}}^{A}\gamma_k p^A - {1\over 3}    
{\bar{n}}^{A}\gamma_k n^A \right]A_k    
- {1\over {2\sqrt {g^2 +{g'}^2}}}\left[    
-(g^2 - {g'}^2 ){\bar{e}}^{A}_{L}\gamma^{\mu} e^{A}_{L}    
+2{g'}^2{\bar{e}}^{A}_{R}\gamma^{\mu} e^{A}_{R} \right. \nonumber \\    
&+& \left.(g^2 + {g'}^2 ){\bar{\nu}}^{A}\gamma^{\mu} \nu^{A}\right]Z_{\mu}-   
{1\over {2\sqrt {g^2 +{g'}^2}}}\left[(g^2 -    
{{g'}^2\over 3} ){\bar{p}}^{A}_{L}\gamma^{\mu} p^{A}_{L}    
-{4{g'}^2\over 3}{\bar{p}}^{A}_{R}\gamma^{\mu} p^{A}_{R}    
-(g^2 + {{g'}^2\over 3} ){\bar{n}}^{A}_{L}\gamma^{\mu} n^{A}_{L}    
\right. \nonumber \\    
&+& \left.{2{g'}^2\over 3}{\bar{n}}^{A}_{R}\gamma^{\mu} n^{A}_{R}    
\right]   Z_{\mu}     
-  {g\over {\sqrt 2}}\left[    
 ({\bar{\nu}}^{A}_{L}\gamma^{\mu} e^{A}_{L} +    
{\bar{p}}^{A}_{L}\gamma^{\mu} U_{AB}n^{B}_{L})W^{+}_{\mu}    
+ \mbox{h.c.} \, \right]\; . \label{eq:22225}    
\end{eqnarray}    
The Hamiltonian, in which we have added the    
primary constraints with the help of Lagrange multipliers    
with their appropriate parities, can be written as  
\newpage   
\begin{eqnarray}    
{\cal H}^{(1)} &=& {\cal H}+    
A_0 \left[\partial_i\Pi_{i}^{A} +    
{igg'\over {\sqrt{ g^2 +{g'}^2}}}    
( \Pi^{W^-}_{i} W^{-}_{i} - \Pi^{W^+}_{i}W^{+}_{i})-    
 {gg'\over {\sqrt {g^2 +{g'}^2}}} \left(    
-{\bar{e}}^{A}\gamma^{0} e^A + {2\over 3}    
{\bar{p}}^{A}\gamma^{0} p^A \right. \right. \nonumber\\    
&-& \left.\left.{1\over 3}    
{\bar{n}}^{A}\gamma^{0} n^A \right) \right]    
+\lambda^Z\Pi^{Z}_{0} +\lambda^{W^+}\Pi^{W^+}_{0}    
+\lambda^{W^-}\Pi^{W^-}_{0}    
+\zeta_{\bar{e}^A}\Pi_{\bar{e}^A}+    
(\Pi_{{e}^A}-i\bar{e}^A \gamma^0)\zeta_{e^A}+    
\zeta_{\bar{\nu}^A}\Pi_{\bar{\nu}^A} \nonumber \\[1ex]    
&+&(\Pi_{{\nu}^A}-i\bar{\nu}^A \gamma^0)\zeta_{\nu^A}+    
\zeta_{\bar{p}^A}\Pi_{\bar{p}^A}+    
(\Pi_{{p}^A}-i\bar{p}^A \gamma^0)\zeta_{p^A}+    
\zeta_{\bar{n}^A}\Pi_{\bar{n}^A}+    
(\Pi_{{n}^A}-i\bar{n}^A \gamma^0)\zeta_{n^A} \;.   
\end{eqnarray}    
The requirement for the primary constraints to be time-independent    
completely defines the odd Lagrange multipliers    
$\zeta_{\bar{e}^A},\ \zeta_{e^A},    
\ \zeta_{\bar{\nu}^A},\ \zeta_{\nu^A},    
\ \zeta_{\bar{p}^A}\ \zeta_{p^A},    
\ \zeta_{\bar{n}^A}\ \zeta_{n^A}$.    
If we substitute the values for the odd Lagrange multipliers one sees    
by direct inspection that    
\begin{equation}    
\left\{ \phi^{(1)}_{A},{\cal H}\right\}_{|\phi^{(1)}=0}=0 \nonumber    
\end{equation}    
is satisfied identically.    
The primary    
constraints for the massive vector fields generate secondary    
constraints   
\begin{eqnarray}\label{eq:22240}   
\phi^{(2)}_{Z} = -& \partial_i&\Pi_{i}^{Z} -    
{ig^2\over {\sqrt{ g^2 +{g'}^2}}}    
( \Pi^{W^-}_{i} W^{-}_{i} - \Pi^{W^+}_{i}W^{+}_{i})    
+M^{2}_{Z} Z_0   \label{eq:222400} \nonumber \\    
&+&{1\over {2\sqrt {g^2 +{g'}^2}}} \left[    
-(g^2 - {g'}^2 ){\bar{e}}^{A}_{L}\gamma^0 e^{A}_{L}    
+2{g'}^2{\bar{e}}^{A}_{R}\gamma^0 e^{A}_{R}    
+(g^2 + {g'}^2 ){\bar{\nu}}^{A}\gamma^0 \nu^{A}\right]  \nonumber \\    
&+&{1\over{2\sqrt{ g^2 +{g'}^2}}}   
\left[(g^2 - {{g'}^2\over 3} ){\bar{p}}^{A}_{L}\gamma^0 p^{A}_{L}    
-{4{g'}^2\over 3}{\bar{p}}^{A}_{R}\gamma^0 p^{A}_{R}    
-(g^2 + {{g'}^2\over 3} ){\bar{n}}^{A}_{L}\gamma^0 n^{A}_{L}    
+{2{g'}^2\over 3}{\bar{n}}^{A}_{R}\gamma^0 n^{A}_{R}    
\right]\;,\\&& \nonumber \\     
\phi^{(2)}_{W^+} = -&\partial_i&\Pi_{i}^{W^+} +    
{igg'\over {\sqrt{ g^2 +{g'}^2}}}    
( \Pi^{A}_{i} W^{-}_{i} - \Pi^{W^+}_{i}A_i)+    
{ig^2\over {\sqrt{ g^2 +{g'}^2}}}    
( \Pi^{Z}_{i} W^{-}_{i} - \Pi^{W^+}_{i}Z_i)    
+M^{2}_{W} W^{-}_{0} \nonumber \\    
&+&{g\over {\sqrt 2}}    
 ({\bar{\nu}}^{A}_{L}\gamma^{0} e^{A}_{L} +    
{\bar{p}}^{A}_{L}\gamma^{0} U_{AB}n^{B}_{L})\;, \\    
&& \nonumber \\    
\phi^{(2)}_{W^-} = -&\partial_i&\Pi_{i}^{W^-} -    
{igg'\over {\sqrt{ g^2 +{g'}^2}}}    
( \Pi^{A}_{i} W^{+}_{i} - \Pi^{W^-}_{i}A_i)-    
{ig^2\over {\sqrt{ g^2 +{g'}^2}}}    
( \Pi^{Z}_{i} W^{+}_{i} - \Pi^{W^-}_{i}Z_i)    
+M^{2}_{W} W^{+}_{0}  \label{eq:222401} \nonumber\\    
&+&{g\over {\sqrt 2}}    
 ({\bar{e}}^{A}_{L}\gamma^{0} \nu^{A}_{L} +    
{\bar{n}}^{A}_{L}U^{*}_{AB}\gamma^{0} p^{B}_{L})\;.    
\end{eqnarray}    
If we impose that the new constraints (\ref{eq:222400})-(\ref{eq:222401})    
are time-independent we find the even Lagrange multipliers    
$\lambda^{W^+},\ \lambda^{W^-},\ \lambda^Z$ and no new constraints    
appear.    
    
We will quantize the proposed model in the Coulomb gauge    
$\partial_i A_i=0$ for the electromagnetic gauge potential.    
We choose sources for the fields, having the relevant  parity.    
The functional integral $Z({\cal J})$ acquires the form    
\begin{equation}\label{eq:22244}    
Z({\cal J}) = \int{ \exp\left [i \int d^4x    
(\pi^{\alpha}q^{\alpha} - {\cal H}+ j^{\alpha}q^{\alpha})  \right]   
\mbox{Sdet}^{1\over 2}\left\{\phi',\phi'\right\}    
\delta (\phi^{(1)})\delta (\phi^{(2)})    
\delta (\partial_i A_i)d\mu(\pi,q) }\;,    
\end{equation}    
where $d\mu(\pi,q) ={\cal D}\pi {\cal D}q$ and    
the sums over $\alpha$ in the exponent of (\ref{eq:22244})    
are given by   
\begin{eqnarray}\label{eq:22245}    
\pi^{\alpha}q^{\alpha} =    
&\Pi^{A}_{i}&\mathaccent 95{A}^i + \Pi^{Z}_{0}\mathaccent 95{Z}^0+    
\Pi^{Z}_{i}\mathaccent 95{Z}^i + \Pi^{W^+}_{0}\mathaccent 95{W}^{+0}+    
\Pi^{W^+}_{i}\mathaccent 95{W}^{+i}+\Pi^{W^-}_{0}\mathaccent 95{W}^{-0}+    
\Pi^{W^-}_{i}\mathaccent 95{W}^{-i}  \nonumber\\[.5ex]     
&+&\Pi_{e^A}\mathaccent 95{e}^{A}+   
\mathaccent 95{\bar{e}}^{A}\Pi_{\bar{e}^A}+    
\Pi_{\nu^A}\mathaccent 95{\nu}^{A}+ \mathaccent    
95{\bar{\nu}}^{A}\Pi_{\bar{\nu}^A}+ \Pi_{p^A}\mathaccent 95{p}^{A}+    
\mathaccent 95{\bar{p}}^{A}\Pi_{\bar{p}^A}+ \Pi_{n^A}\mathaccent    
95{n}^{A}+ \mathaccent 95{\bar{n}}^{A}\Pi_{\bar{n}^A}\;,  \\    
& & \nonumber \\    
j^{\alpha}q^{\alpha} =    
&J^{\mu}_{A}&A_{\mu}+J^{\mu}_{Z}Z_{\mu}+    
J^{\mu}_{-}W^{+}_{\mu}+J^{\mu}_{+}W^{-}_{\mu}    
+\bar{e}^A\eta_{\bar{e}^A}+\bar{\eta}_{e^A}e^A \nonumber \\[.5ex]    
&+&\bar{\nu}^A\eta_{\bar{\nu}^A}+\bar{\eta}_{\nu^A}\nu^A+    
\bar{p}^A\eta_{\bar{p}^A}+\bar{\eta}_{p^A}p^A+    
\bar{n}^A\eta_{\bar{n}^A}+\bar{\eta}_{n^A}n^A\;.   
\end{eqnarray}    
By direct inspection one sees that the superdeterminant,    
restricted to the constraints by the $\delta$-functions \cite{Git}    
is independent of the fields and reduces    
to a constant multiplier of $Z({\cal J})$.    
    
Given the integral representation of the $\delta$-function,    
one may place the constraints    
$\phi^{(1)}_{A},\ \phi^{(2)}_{W^+},    
\ \phi^{(2)}_{W^-}$ and $\phi^{(2)}_{Z}$ in the exponent of    
the functional integral. As a result we have additional integration    
over the variables $A_0,\ \lambda^{W^+},\ \lambda^{W^-}$    
and $\lambda^Z$.    
A change of variables, which makes the corresponding integrals    
of a Gaussian type, reads    
\begin{equation}    
W^{+}_{0} + \lambda^{W^+} \rightarrow W^{+}_{0}\;,\qquad    
W^{-}_{0} + \lambda^{W^-} \rightarrow W^{-}_{0}\;,\qquad    
Z_{0} + \lambda^{Z} \rightarrow Z_{0}\;.    
\end{equation}    
These integrals do not contribute to the {\em normalized}    
$Z({\cal J})$.    
In the next step one has to take the integrals over    
$\Pi^{A}_{i},\ \Pi^{Z}_{i},\ \Pi^{W^+}_{i}$ and $\Pi^{W^-}_{i}$.    
A change of variables, that separates the integrations, is    
\begin{eqnarray}\label{eq:22246}    
&&\Pi^{A}_{i}\rightarrow \Pi^{A}_{i} -    
\left(  \mathaccent 95{A}^i+    
\partial_i A_0 +  {igg'\over {\sqrt{ g^2 +{g'}^2}}}    
\left( W^{+}_{0} W^{-}_{i} - W^{+}_{i} W^{-}_{0}\right)\right)\;,\\    
&&{\Pi}_{i}^{Z} \rightarrow {\Pi}_{i}^{Z}  -\left(    
\mathaccent 95{Z}^i+    
\partial_i Z_0 +  {ig^2\over {\sqrt{ g^2 +{g'}^2}}}    
\left( W^{+}_{0} W^{-}_{i} - W^{+}_{i} W^{-}_{0}\right)\right)\;, \\    
&&{\Pi}_{i}^{W^+} \rightarrow \Pi^{W^+}_{i}-    
\left(\mathaccent 95{W}^{-i}+    
\partial_i W_{0}^{-} +  {igg'\over {\sqrt{ g^2 +{g'}^2}}}    
\left( W^{-}_{0} A_{i} - A_0 W^{-}_{i} \right) +    
  {ig^2 \over {\sqrt{ g^2 +{g'}^2}}}    
\left( W^{-}_{0} Z_{i} - Z_0 W^{-}_{i} \right)\right)\;,\\    
&&{\Pi}_{i}^{W^-} \rightarrow \Pi^{W^-}_{i} -    
\left(  \mathaccent 95{W}^{+i}+    
\partial_i W_{0}^{+} +  {igg'\over {\sqrt{ g^2 +{g'}^2}}}    
\left( A_0 W^{+}_{i} -  W^{+}_{0}A_i \right) +    
  {ig^2 \over {\sqrt{ g^2 +{g'}^2}}}    
\left( Z_0 W^{+}_{i} -  W^{+}_{0} Z_i \right)\right)\;.    
\end{eqnarray}    
    
For the normalized functional integral one obtains finally   
\begin{eqnarray}\label{eq:22249}    
{Z({\cal J})\over Z_0({\cal J})} &= &\int{ \exp \left[ i\int {d^4x    
({\cal L}+ j^{\alpha}q^{\alpha})}\right]    
\delta (\partial_i A_i)d\mu(q)}\;.    
\end{eqnarray}    
\section{A Ward Identity}    
    
The functional integral (\ref{eq:22249}) can be written in the    
form:    
\begin{eqnarray}\label{eq:22260}    
{Z({\cal J})\over Z_0({\cal J})} = \int{ \exp \left[ i\int {d^4x    
\left({\cal L}_{(0)} + {\cal L}_{(\mbox{\scriptsize g.f.})}+    
{\cal L}_{(\mbox{\scriptsize src.})}    
+ {\cal L}_{(\mbox{\scriptsize int.})}\right)}\right]    
d\mu (q)}\;,    
\end{eqnarray}    
where the terms in the exponent are the Lagrangian of    
the free theory, the gauge fixing part    
${\cal L}_{(\mbox{\scriptsize g.f.})} =   
-{1\over 2}(\partial_{\mu}A^{\mu})^2$    
for the Feynman gauge of $A_{\mu}$, the source part and the    
interaction. One easily checks    
that all nonzero vertices (i.e. those without Higgs lines),     
as well as the 2-point free Green's   
functions for the photon, vector-meson and spinor fields,    
coincide with the corresponding ones from the WS model ~\cite{Lee} in the   
unitary gauge.   
The physical consequences of the model should not    
depend on the  gauge transformations. Introducing the restriction    
that the functional integral $Z({\cal J})$ is    
gauge-invariant, one finds an equation in variational    
derivatives, which represents the Ward identity.    
The infinitesimal gauge transformations from the    
 electromagnetic gauge subgroup \cite{R} are    
\begin{eqnarray}\label{eq:22290}    
&&{A'}_{\mu} = A_{\mu} + {1 \over e} \partial_{\mu} \alpha\;, \qquad    
\quad {Z'}_{\mu} = Z_{\mu}\;,\label{eq:22291} \\    
&&{W'}^{+}_{\mu} = W^{+}_{\mu} +i \alpha W^{+}_{\mu}\;, \qquad    
{W'}^{-}_{\mu} = W^{-}_{\mu} -i \alpha W^{-}_{\mu}\;, \\    
&&{e'}^{A} = e^{A} -i \alpha e^{A}\;, \qquad    
\quad \quad {\bar{e'}}^{A} = \bar{e}^{A} +i \alpha \bar{e}^{A}\;, \\    
&&{\nu'}^{A} = \nu^{A} \;, \qquad   \qquad    
\qquad \quad {\bar{\nu'}}^{A} = \bar{\nu}^{A}\; , \\    
&&{p'}^{A} = p^{A} +{2i\over 3} \alpha p^{A}\;, \qquad    
\quad {\bar{p'}}^{A} = \bar{p}^{A} -{2i\over 3} \alpha \bar{p}^{A}\;, \\    
&&{n'}^{A} = n^{A} -{i\over 3} \alpha n^{A}\;, \qquad    
\quad {\bar{n'}}^{A} = \bar{n}^{A} +{i\over 3}    
\alpha \bar{n}^{A}\;.\label{eq:22292}    
\end{eqnarray}    
The transformations (\ref{eq:22291})-(\ref{eq:22292}) result in    
an additional exponential part in the functional integral    
\begin{eqnarray}\label{eq:22293}    
{Z^{(0)}({\cal J})\over Z_0({\cal J})} &=& \int \exp \left[i\int\ d^4x    
\left(-{1\over e}\partial_{\mu}A^{\mu}    
\partial_\nu \partial^\nu\alpha +    
{1\over e} J^{\mu}_{A}\partial_{\mu}\alpha +    
i(J^{\mu}_{-}W^{+}_{\mu}-J^{\mu}_{+}W^{-}_{\mu})\alpha    
\right.\right. \nonumber \\    
&& \quad \left.\left. +i(\bar{e}^A\eta_{\bar{e}^A}-   
\bar{\eta}_{e^A}e^A)\alpha -    
{2i\over 3}(\bar{p}^A\eta_{\bar{p}^A}-\bar{\eta}_{p^A}p^A)\alpha+    
{i\over 3}(\bar{n}^A\eta_{\bar{n}^A}-\bar{\eta}_{n^A}n^A)\alpha   
\right)\right] \times    
\nonumber \\    
&& \qquad\qquad\qquad\qquad\qquad\qquad\qquad   
\qquad\qquad\qquad   
\times \exp \left[ i\int { d^4x{\cal L}_{(\mbox{\scriptsize eff.})}}   
\right]   d\mu (q)\;.    
 \end{eqnarray}    
If we expand the exponent over the infinitesimal parameter    
$\alpha$ and substitute the fields with their variational    
derivatives up to first    
order in the expansion we get   
\begin{eqnarray}\label{eq:222100}    
&&\left[-i \,  
\partial_\nu \partial^\nu\,\partial_{\mu}   
{\delta \over \delta J_{A_{\mu}}} +    
\partial_{\mu}J^{\mu}_{A} -    
e\left({\delta \over \delta_{\mu}J_{-\mu}}J^{\mu}_{-} -    
J^{\mu}_{+}{\delta \over \delta J_{+\mu}}\right)+    
e\left(\bar{\eta}_{e^A}{\delta \over \delta \bar{\eta}_{e^A}} +    
{\delta \over \delta\eta_{\bar{e}^A}}\eta_{\bar{e}^A}\right)   
\right. \nonumber \\    
&&\quad \left. -{2\over 3}e\left(\bar{\eta}_{p^A}   
{\delta \over \delta \bar{\eta}_{p^A}} +    
{\delta \over \delta\eta_{\bar{p}^A}}\eta_{\bar{p}^A}\right)+    
{1\over 3}e\left(\bar{\eta}_{n^A}{\delta \over \delta \bar{\eta}_{n^A}} +    
{\delta \over \delta\eta_{\bar{n}^A}}\eta_{\bar{n}^A}\right)\right]    
Z({\cal J})=0\;, \label{eq:222110}    
\end{eqnarray}    
where the variational derivatives act directly on $Z({\cal J})$.  
With the transformation    
\begin{eqnarray}   
Z({\cal J})=e^{iW({\cal J})}\;,    
\end{eqnarray}    
Eq. (\ref{eq:222100}) acquires the form    
\begin{eqnarray}\label{eq:222111}    
&&\partial_{\nu}\partial^{\nu}\,   
 \partial_{\mu}{\delta W \over \delta J_{A\mu}} +    
\partial_{\mu}J^{\mu}_{A}    
-ie\left({\delta W \over \delta_{\mu}J_{-\mu}}J^{\mu}_{-} -    
J^{\mu}_{+}{\delta W\over \delta J_{+\mu}}\right)+    
ie\left(\bar{\eta}_{e^A}{\delta W\over \delta \bar{\eta}_{e^A}} +    
{\delta W\over \delta\eta_{\bar{e}^A}}\eta_{\bar{e}^A}\right)   
\nonumber \\[.5ex]    
&&\quad    
-{2i\over 3}e\left(\bar{\eta}_{p^A}{\delta W\over \delta \bar{\eta}_{p^A}} +   
{\delta W\over \delta\eta_{\bar{p}^A}}\eta_{\bar{p}^A}\right)+    
{i\over 3}e\left(\bar{\eta}_{n^A}{\delta W\over \delta \bar{\eta}_{n^A}} +    
{\delta W\over \delta\eta_{\bar{n}^A}}\eta_{\bar{n}^A}\right)=0\;.    
\end{eqnarray}    
It is convenient to rewrite (\ref{eq:222111}) as an equation for the    
vertex function    
\begin{eqnarray}\label{eq:222112}    
\Gamma = W({\cal J}) - \int d^4x \, j^{\alpha}q^{\alpha}\;.    
\end{eqnarray}    
    
We may express the sources through variational derivatives over    
the fields    
\begin{eqnarray}\label{eq:222113}    
&&J^{\mu}_{A}\ \rightarrow \  -{\delta\Gamma \over \delta A_{\mu}}   
\; , \qquad \qquad  \  
J^{\mu}_{Z}\  \rightarrow \  -{\delta\Gamma \over \delta Z_{\mu}}, \\    
&&J^{\mu}_{+}\  \rightarrow \  -{\delta\Gamma \over \delta W_{\mu}^{-}}   
\;, \qquad \qquad    
J^{\mu}_{-} \ \rightarrow \  -{\delta\Gamma \over \delta W_{\mu}^{+}}\;, \\    
&&\eta_{\bar{e}^A}\  \rightarrow \  -{\delta\Gamma \over \delta    
\bar{e}^{A}}\;,\qquad \qquad \ \    
\bar{\eta}_{{e}^A}\  \rightarrow \  {\delta\Gamma \over    
\delta {e}^{A}}\;,\\    
&&\eta_{\bar{\nu}^A}\  \rightarrow \ -{\delta\Gamma \over    
\delta \bar{\nu}^{A}}\;, \qquad \qquad \  \   
\bar{\eta}_{{\nu}^A} \rightarrow \    
{\delta\Gamma \over \delta {\nu}^{A}}\;,\\    
&&\eta_{\bar{p}^A}\  \rightarrow \    
-{\delta\Gamma \over \delta \bar{p}^{A}}\;, \qquad \qquad \ \   
\bar{\eta}_{{p}^A}    
\ \rightarrow\  {\delta\Gamma \over \delta {p}^{A}}\;,\\    
&&\eta_{\bar{n}^A}    
\ \rightarrow \ -{\delta\Gamma \over \delta \bar{n}^{A}}\;,    
\qquad \qquad \  \   
\bar{\eta}_{{n}^A} \ \rightarrow \ {\delta\Gamma \over \delta    
{n}^{A}}\;.\\    
\end{eqnarray}    
If we recall the standard variational derivative expressions for the   
fields in the path integral formulation (\ref{eq:22249})  and take into account    
the definition of the vertex function (\ref{eq:222112}) we find   
\begin{eqnarray}\label{eq:222115}    
&&{\delta W \over \delta J^{\mu}_{A}}\  \rightarrow\  A_{\mu}\;,\qquad \qquad    
\ \ {\delta W \over \delta J^{\mu}_{Z}}\  \rightarrow\  Z_{\mu}\;,\\    
&&{\delta W \over \delta J^{\mu}_{+}}\  \rightarrow\  W_{\mu}^{-}\;,\qquad \qquad    
\ {\delta W \over \delta J^{\mu}_{-}}\  \rightarrow \ W_{\mu}^{+}\;,\\    
&&{\delta W \over \delta \bar{\eta}_{e^A}}\  \rightarrow\  e^A\;,\qquad \qquad    
-{\delta W \over \delta \eta_{\bar{e}^A}}\  \rightarrow \ \bar{e}^A\;,\\    
&&{\delta W \over \delta \bar{\eta}_{\nu^A}}\  \rightarrow\  \nu^A\;,\qquad \qquad    
-{\delta W \over \delta \eta_{\bar{\nu}^A}}\  \rightarrow\  \bar{\nu}^A\;,\\    
&&{\delta W \over \delta \bar{\eta}_{p^A}}\  \rightarrow\  p^A\;,\qquad \qquad    
-{\delta W \over \delta \eta_{\bar{p}^A}}\  \rightarrow \ \bar{p}^A\;,\\    
&&{\delta W \over \delta \bar{\eta}_{n^A}}\  \rightarrow\  n^A,\qquad \qquad    
-{\delta W \over \delta \eta_{\bar{n}^A}}\  \rightarrow \ \bar{n}^A\;.    
\end{eqnarray}    
Finally using (\ref{eq:222111}) we derive   
\begin{eqnarray}\label{eq:2222001}    
&& \partial_{\nu}\partial^{\nu}\,\partial_{\mu}A^{\mu} -    
\partial_{\mu}{\delta \Gamma \over \delta A_{\mu}}-    
ie\left({\delta \Gamma \over \delta W_{\mu}^{-}}W_{\mu}^{-} -    
W_{\mu}^{+}{\delta \Gamma \over \delta W_{\mu}^{-}}\right)+    
ie\left(\bar{e}^A{\delta \Gamma \over \delta \bar{e}^A} +   
{\delta \Gamma \over \delta e^A}e^A\right) \nonumber \\[.5ex]    
&&\quad -{2i\over 3}e\left(\bar{p}^A{\delta \Gamma \over \delta \bar{p}^A} +    
{\delta \Gamma \over \delta p^A}p^A\right)+    
{i\over 3}e\left(\bar{n}^A{\delta \Gamma \over \delta \bar{n}^A} +    
{\delta \Gamma \over \delta n^A}n^A\right) =0\;.    
\end{eqnarray}    
Equation (\ref{eq:2222001}) gives relations between the Green's functions    
of the charged fields and the electromagnetic gauge potential.    
Taking variational derivatives gives the Ward identities for    
the electromagnetic interaction of the    
fields in the proposed model.    
    
If we substitute the action   
\begin{eqnarray}\label{eq:2222112}    
{\cal S}=\int\ {d^4x \,{\cal L}}    
\end{eqnarray}    
with a suitably  regularized gauge-invariant  action    
\begin{eqnarray}\label{eq:2222113}    
{\cal S}_{\Lambda}=\int\ {d^4x \,{\cal L}_{\Lambda}} \;,   
\end{eqnarray}    
($\Lambda$ being a regularizing parameter) we may    
derive analogous equation, which gives     
relations among    
the regularized Green's functions in arbitrary order in    
perturbation theory.    

\vskip 1cm     
\begin{center}     
{\bf ACKNOWLEDGMENTS}     
\end{center}     
\vskip 0.25cm     
One of us (V.R.) is grateful to Prof. R. Kerner and the     
Laboratoire de Physique des Particules, Universit\'e Pierre et     
Marie Curie, Paris, for their hospitality during a visit when      
this work was started. This work was partly supported (I.V.) by the    
DOE Research Grant under Contract No. De-FG-02-93ER-40764.


\newpage     
   
   
\begin{thebibliography}{99}    
    
\bibitem{We}{  S.~Weinberg, {\em Phys. Rev. Lett.},    
{\bf 19}, (1967) 1264 .  }    
%
\bibitem{Sa}{ A.~Salam, {In: \em Elementary Particle Physics: Relativistic    
Groups and Analyticity, Nobel Symposium No 8 \\   
(Ed. N.Svartholm, Almqvist and    
Wiksell, Stockholm, 1968) 367.} }    
%
\bibitem{Hi}{ P.~W.~Higgs, {\em Phys. Rev. Lett.},    
{\bf 12}, (1964) 132.  }    
%
\bibitem{Ab}{ E.~Abers and B.~Lee: {\em Phys. Rep.},    
{\bf 9C} No. 1(1973) 2. }    
%
\bibitem{La}{ J.~LaChapelle, {\em J. Math. Phys.},    
{\bf 35} (1994) 2199.}    
%
\bibitem{Pa}{ M.~Pawlowski and R.~Raczka,   
{\em Found. Phys.},{\bf 24} (1994) 1305.}   
%
\bibitem{R}{V.~Rizov and I.~Vitev, {\em A Note on the Gauge Group of the   
Electroweak Interactions,} hep-ph/0009169.}   
%
\bibitem{Ro}{G.~Roepstorff and Ch.~Vehns,    
{\em Towards a Unified Theory of Gauge    
and Yukawa Interactions,} hep-ph/0006065.}   
%
\bibitem{Lee}{L.~F.~Lee and T.~P.~Cheng, {\em ``Gauge    
Theory of Elementary Particle Physics",}    
Clarendon Press, Oxford, 1984.}    
%
\bibitem{Sen}{P.~Senjanovic, {\em Ann. of Phys.},    
{\bf 100} (1976) 227}    
%
\bibitem{Git}{D.~M.~Gitman and I.~V.~Tyutin, {\em ``Quantization   
of  Fields with Constraints"}, Springer-Verlag,    
Berlin Heidelberg, 1990.}    
%
\bibitem{Sl}{L.~D.~Faddeev and A.~A.~Slavnov,   
{\em ``Gauge Fields. An Introduction to Quantum Theory"},     
Addison-Wesley Publishing Company, Redwood City, California, 1991.}    
%
\end{thebibliography}
\end{document}